\documentclass[a4paper,twoside,12pt]{article}
\usepackage[dvips]{graphicx}
\usepackage{epsfig}

\input{psfig.sty}
\newcommand{\eg}{{\em e.g.}}
\newcommand{\ie}{{\em i.e.}}

\newcommand{\ccbar}{$c\bar{c}$}

\newcommand{\lepto}{{\sc Lepto}}
\newcommand{\pythia}{{\sc Pythia}}

\newcommand{\xf}{$x_{F} \,$}
\newcommand{\pt}{$p_{\perp} \,$}

\newcommand{\etal}{{\it et al.}}
\newcommand{\jpsi}{$J/\psi$}
\newcommand{\psip}{${\psi}^{\prime}$}
\newcommand{\Y}{$\Upsilon \,$}

\def\ccb{$c\overline{c} \,\,$}

\def\mccb{m_{c\overline{c}}}

\def\nlo{{\sc CEM-NLO}}

\def\sci{{\sc SCI-PYTHIA}}

\def\be{\begin{equation}}
\def\ee{\end{equation}}
\def\bea{\begin{eqnarray}}
\def\eea{\end{eqnarray}}

\def\ccb{$c\overline{c} \,\,$}

\def\lsim{\mathrel{\rlap{\lower4pt\hbox{\hskip1pt$\sim$}}
    \raise1pt\hbox{$<$}}}         
\def\gsim{\mathrel{\rlap{\lower4pt\hbox{\hskip1pt$\sim$}}
    \raise1pt\hbox{$>$}}}         
\begin{document}
\titlepage

\noindent
TSL/ISV-2001-0256 \\
November 2001       
\vspace*{10mm}
\begin{center}
  \begin{Large}
  \begin{bf}
Soft and hard QCD dynamics\\
in hadroproduction of charmonium\\
  \vspace*{5mm}
  \end{bf}
  \end{Large}
  \begin{large}
    C.~Brenner~Mariotto$^{ab}$, M.B. Gay Ducati$^{b}$, G. Ingelman$^{ac}$\\  
  \end{large}
\end{center}
$^a$~High Energy Physics, Uppsala University, Box 535, S-75121 Uppsala, Sweden\\
$^b$~Institute of Physics, Univ.\ Fed.\ do Rio Grande do Sul, 
Box 15051, CEP 91501-960 Porto Alegre, Brazil\\
$^c$~Deutsches~Elektronen-Synchrotron~DESY, Notkestrasse~85,~D-22603~Hamburg,~Germany\\
\vspace*{5mm}
\begin{quotation}
\noindent
{\bf Abstract:}
Both hard and soft QCD dynamics are important in charmonium production, as presented here through a next-to-leading order QCD matrix element calculation combined with the colour evaporation model. Observed $x_F$ and $p_\perp$ distributions of $J/\psi$ in hadroproduction at fixed target and $p\bar{p}$ collider energies are reproduced. Quite similar results can also be obtained in a more phenomenologically useful Monte Carlo event generator where the perturbative production of \ccbar \ pairs is instead obtained through leading order matrix elements and the parton shower approximation of the higher order processes. The soft dynamics may alternatively be described by the soft colour interaction model, originally introduced in connection with rapidity gaps. We also discuss the relative rates of different charmonium states and introduce an improved model for mapping the continuous \ccbar \ mass spectrum on the physical charmonium resonances. 
\end{quotation}

\section{Introduction}
The interplay between perturbative  QCD (pQCD) and non-perturbative QCD is particularly important in the production of charmonium states in hadronic interactions. The charm quark mass $m_c$ provides a scale which is large enough to make $\alpha_s(m_c^2)$ sufficiently small for the application of pQCD to describe the parton level production of a \ccbar \ pair. The subsequent formation of a bound state charmonium state, on the other hand, is a non-perturbative phenomenon where important new effects enter. In the pQCD-based
Colour Singlet Model~\cite{CSM} (CSM), the predicted cross section turned out to be more than one order of magnitude lower than what has been observed for high-$p_\perp$ charmonium at the Tevatron~\cite{high-pt-data,D0}. However, these data can be well described by the Colour Evaporation Model (CEM) \cite{HALZENquantit}, the Soft Colour Interaction model (SCI) \cite{sci-onium} and the Colour Octet Model (COM)~\cite{COM}, where the perturbatively produced $c\bar{c}$ pair in a colour octet state can be transformed into a colour singlet state through soft interactions which can be viewed as non-perturbative gluon exchange. These models make different explicit attempts to describe the soft QCD dynamics which increases the cross section for charmonium production by more than one order of magnitude. 

In this paper we present a detailed study of charmonium production within these 
models. For the perturbative \ccbar \ production we employ exact matrix elements which are available up to next-to-leading order \cite{NLOsi,NLO}. As an alternative we use leading order matrix elements complemented with parton showers giving an approximate treatment of arbitrarily high orders. This latter alternative is used in Monte Carlo programs which also include non-perturbative hadronisation resulting in the simulation of complete events.  These event generators are phenomenologically very useful, since any physical observable can be extracted and compared in detail with experimental results. Our finding that this approach can effectively give the same result as the next-to-leading order calculation is therefore comforting. 

Based on these perturbative \ccbar \ production mechanisms, we use the CEM and SCI models to form charmonium states. Here, one has to consider how the continuous \ccbar \ mass spectrum should be mapped on the discrete spectrum of physical charmonium resonances. Previously used schemes have employed simple parameters that are fitted to data or factors from spin statistics. The latter approach is here developed further in order to introduce a correlation between the invariant mass of the \ccbar \ pair and the masses of the different charmonium states. This improves the description of the observed relative rates of, \eg ,  \jpsi \ and \psip. 

The models are tested in detail against data on \jpsi \ production in hadronic interactions, which are available from fixed target and collider experiments covering a large energy range. Total cross sections and differential distributions in $x_F$ and $p_\perp$ can, in general, be well described. This gives, in particular, additional insights about important soft effects and should thereby help to obtain a better understanding of non-perturbative QCD.

\section{Theoretical description of charmonium production}
The theoretical description separates the hard and soft parts of the process based on the factorisation theorem in QCD. Thus, we will first consider the perturbative production at the parton level, then the non-perturbative formation of bound charmonium states and finally their combination into complete models.

\begin{figure}[thbp]
\begin{center}
\mbox{
\epsfig{file=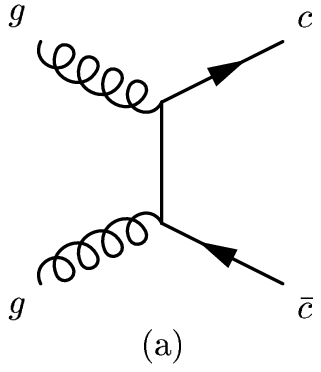,width=4cm}
\hspace{10mm}
\epsfig{file=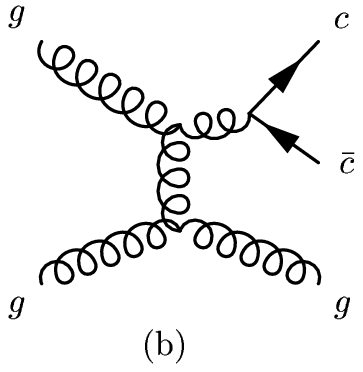,width=4cm}
}
\vspace{0.2cm}
\caption{Illustration of $c\bar{c}$ production processes in (a) leading order and (b) next-to-leading order perturbative QCD.}
\label{fig:feyn}
\end{center}
\end{figure}

\subsection{Production of \ccbar \ in perturbative QCD}
Perturbative QCD should be applicable for \ccbar \ production, since the charm quark mass $m_c$ is large enough to make $\alpha_s(m_c^2)$ a small expansion parameter. The leading order (LO) processes are $gg \rightarrow c\bar{c}$ 
(Fig.~\ref{fig:feyn}a) and $q\bar{q} \rightarrow c\bar{c}$. Heavy quark 
production is, however, known to have large contributions from next-to-leading 
order (NLO) diagrams \cite{NLOsi}. Virtual corrections to the leading order 
processes together with soft and collinear gluon emissions give an increase 
of the cross section, which can be approximately described as an overall 
$K$-factor multiplying the leading order cross-section. In addition, some 
NLO tree diagrams with a third hard parton contribute significantly, with the most important one  shown in Fig.~1b. Although this is an ${\cal O}(\alpha_s^3)$ process with a gluon splitting $g\to c\bar{c}$ added to the basic $2\to 2$ process $gg\to gg$, it is numerically large since $gg\to gg$ has a much larger ${\cal O}(\alpha_s^2)$ cross section than the LO $c\bar{c}$ production processes. 

In order to treat all these contributions properly, one must use the available 
NLO matrix element calculations with explicit non-zero quark masses \cite{NLOsi,
NLO}. This is particularly important at smaller energies where the correct threshold behaviour of the cross section is vital. We therefore use the available computer program \cite{NLO} which calculates total and differential cross sections based on the complete set of diagrams that contribute up to NLO, \ie \ ${\cal O}(\alpha_s^3)$, in the $\overline{MS}$ scheme. It is here possible to separate the LO ${\cal O}(\alpha_s^2)$ contribution, such that the effect of the NLO corrections can be seen. The free parameters of the NLO calculation are the charm quark mass $m_c$, and the factorisation and renormalization scales. The latter are taken as their default options in the program, \ie \ the transverse mass squared $\mu^2=m_c^2 + \langle p_\perp^2 \rangle$ where $\langle p_\perp^2 \rangle = (p_{c\; \perp}^2+p_{\bar{c}\; \perp}^2)/2$ is the average transverse momentum squared of the charm and anti-charm quarks. The charm quark mass is the main parameter that affects the normalization of the cross section. The value  $m_c=1.5$~GeV is found to work well in our comparisons with data below. 

The LO processes alone or with NLO virtual corrections together with soft and collinear emissions, cannot produce a $J/\psi$ at high-\pt . The reason is that the essentially zero \pt \ of the initial partons cannot be conserved with the \ccbar \ pair at high \pt \ unless there is another hard parton to balance its \pt . Of importance for high-\pt \ $J/\psi$ production are therefore NLO tree diagrams having a third hard parton. The dominating NLO contribution is the diagram in Fig.~1b. At the high energies available at colliders, still higher orders are expected to be important,  since many gluons can be emitted and their virtuality need not be very large in order to allow a split into a $c\bar{c}$ pair. Higher orders than NLO are not available in terms of exact matrix elements, but can be approximately described using the parton shower (PS) approach available in some Monte Carlo event generators. 

As an alternative method to describe the pQCD production of \ccbar \ pairs we therefore use the \pythia \ \cite{Pythia} Monte Carlo, where all QCD $2\to 2$ processes are included with their corresponding matrix elements and the incoming and outgoing partons may branch as described by the DGLAP equations \cite{dglap}. A \ccbar \ pair can then be produced as described by the LO matrix elements for $q\bar{q} \rightarrow c\bar{c}$ and $gg \rightarrow c\bar{c}$ (with explicit $m_c$ dependence) or in a gluon splitting $g \rightarrow c\bar{c}$ in the parton shower. Although the latter is an approximation, it has the advantage of covering arbitrarily high orders and not be limited to NLO. 
We use the default in \pythia \ for the factorisation and renormalization scales, \ie \ the transverse mass squared of the emerging partons in the $2\to 2$ processes (giving the same scales as in the NLO program). The charm mass, which affects the normalization of the cross section, is chosen as $m_c=1.35$~GeV to get the overall agreement with data to be shown below. 

The Monte Carlo approach gives important phenomenological advantages. Models for soft QCD processes can be added resulting in the generation of complete events that can be analysed in the same way as real events. Thus, any experimental observable can be extracted, including all sorts of cuts or requirements, such that a detailed comparison with data can be made. The NLO matrix elements, which contain corrections that cannot easily be interpreted in probabilistic terms, 
have not been implemented in a Monte Carlo program. Since both these approaches to describe pQCD production of \ccbar \ pairs have their advantages and disadvantages, one should investigate to what extent they agree and can be used for various purposes. 

\subsection{Charmonium formation in non-perturbative QCD}
The formation of bound hadron states occurs through processes with small momentum transfers such that the corresponding $\alpha_s$ is large and prevents the use of perturbation theory. The lack of an appropriate method to calculate non-perturbative processes, forces us to use phenomenological models to describe the formation of charmonium states from perturbatively produced \ccbar \ pairs. As mentioned in the Introduction, there are a few models that may do this based on the general idea that the colour charge of a \ccbar \ pair can be changed from octet to singlet through soft QCD processes. A colour singlet \ccbar \ pair with an invariant mass below the threshold for open charm ($m_{c\bar{c}}<2m_D$) will then form a charmonium state. In this way, the more abundantly produced colour octet \ccbar \ pairs will contribute and increase the charmonium production by large factors. Two models of this kind have been investigated and are now described. 

\subsubsection{The colour evaporation model}
The basic hypothesis in the CEM model (\cite{HALZENquantit} and references therein) is that the colour exchange in the soft interactions randomise the colour charges such that no information remains of the colour configuration given by the preceeding hard interactions. Probabilities for colour charge states can then instead be obtained from colour SU(3) algebra, with the relation $3\otimes \bar{3}=1\oplus 8$ being applicable to a \ccbar \ pair composed of a triplet and an antitriplet. With all colour charge states having equal weight, this implies that the \ccbar \ will have a probability $1/9$ to be in a colour singlet state and $8/9$ to be in a colour octet state. It is then assumed that all colour singlet \ccbar \ pairs with 
invariant mass below the threshold for open charm will form a charmonium state. Colour singlet states above this threshold, as well as the \ccbar \ pairs in a colour octet state (independently of invariant mass) will produce open charm through the hadronisation mechanism. The cross section for charmonium and open charm production can then be written as
\begin{eqnarray}
\sigma_{charmonium}&=& \frac{1}{9}\int_{2m_c}^{2m_D}dm_{c\overline{c}}
	\frac{d\sigma_{c\overline{c}}}{dm_{c\overline{c}}} 
\label{1} \\
\sigma_{open}&=& \frac{8}{9}\int_{2m_c}^{2m_D}
dm_{c\overline{c}}
	\frac{d\sigma_{c\overline{c}}}{dm_{c\overline{c}}}
	+ \int_{2m_D}^{\sqrt s}dm_{c\overline{c}}
	\frac{d\sigma_{c\overline{c}}}
{dm_{c\overline{c}}} \,\,,
\label{2}
\end{eqnarray}
where $m_{c\bar{c}}$ is the invariant mass of the \ccbar \ pair, $m_c$ is the charm quark mass and $2m_D$ is the $D\overline{D}$ threshold. The differential parton level cross section  $d\sigma_{c\overline{c}}/dm_{c\overline{c}}$ is the usual convolution of the perturbative QCD cross section with the parton density functions for the initial hadrons. 

This total charmonium cross section is then split on the different charmonium states 
\begin{eqnarray}
\sigma_i=\rho_i\,\sigma_{charmonium}\,\,\,\,\,\,  
(i=J/\psi, \eta_c, \chi_c, \psi^{\prime},...) \,.
\label{3}
\end{eqnarray} 
where the relative rates $\rho_{i}$ are assumed to be independent of process and energy. These non-perturbative parameters must in practice be determined from comparison with data. For \jpsi, this fraction is found to be in the range $\rho_{J/\psi}=$0.4--0.5 \cite{HALZENquantit}.

The CEM is one of the simplest approaches to colour neutralization, where the effect of soft  interactions is implicit in the non-perturbative factors. It gives a good phenomenological description of several production processes up to the present energy domain \cite{satz,schuler,HALZENquantit,ZdecaysGreg}. 
CEM also provides an alternative to Pomeron-based models for the description of photoproduction of $J/\psi$, $D$ and $\Upsilon$ at HERA \cite{CEMnosso}. 

\subsubsection{The soft colour interaction model}
The SCI model \cite{sci} introduces a more explicit description of the soft interactions between partons and hadron remnants emerging from a hard scattering process. In these soft processes, the small momentum transfers are 
not important and can be neglected, at least in a first approximation. Instead, it is the colour exchanges that are important, since they change the colour topology of the event and thereby affect the hadronic final state. The model introduces an explicit mechanism where colour-anticolour, corresponding to non-perturbative gluons, can be exchanged between partons and remnants. This can be viewed as the partons interacting softly with the colour medium, or colour background field, of the initial hadron as they propagate through it. This should be a natural part of the process in which `bare' partons are `dressed' into non-perturbative ones and the confining colour flux tube between them is formed. These colour exchanges lead to different topologies of the confining colour force fields (strings) and thereby to different hadronic final states after hadronisation. 

The SCI model is implemented in the Lund Monte Carlo programs \lepto \ \cite{Lepto} for deep inelastic scattering and \pythia \ \cite{Pythia} for hadron-hadron collisions. The hard parton level interactions are given by standard perturbative matrix elements and parton showers, which are not altered by softer non-perturbative effects. The probability to exchange such a soft gluon within a pair of partons and remnants cannot be calculated and is therefore taken to be a constant given by a phenomenological parameter $R$, which is the only free parameter of the model. As a result of these exchanges, the colour connections, in terms of strings between partons, are changed before the standard Lund model \cite{lund} is applied for hadronisation.  

The SCI model gives a novel explanation of rapidity gap events in both deep inelastic scattering \cite{sci} and hard processes in $p\bar{p}$ at the Tevatron \cite{gaps}, always using the same value $R=0.5$ of the SCI parameter. Applying the same Monte Carlo implementation, it was found that the Tevatron data on 
high-$p_\perp$ charmonium and bottomonium are also well reproduced \cite{sci-onium}. The increased production rate is here given by the possibility for a perturbatively produced \ccbar \  pair in a colour 
octet state to be transformed into a singlet state as a result of these soft 
colour interactions. The mapping of $c\bar{c}$ pairs, with mass below the 
threshold for open charm production, is here made based on spin statistics 
which avoids introducing further free parameters. This results in a fraction of a specific quarkonium state $i$ with total angular momentum $J_i$ given by 
\begin{equation}
f_i = \frac{\Gamma_i}{\sum_k \Gamma_k} 
\end{equation}
where $\Gamma = (2J_i+1)/n_i$ corresponds to a partial width. Radially excited states are here suppressed through their main quantum number $n_i$. This model was found to give a correct description of the different heavy quarkonium states observed at the Tevatron \cite{sci-onium}. 

\subsection{Forming the complete models}
In principle one could add the CEM or SCI models for the soft processes to any of the descriptions for the hard pQCD processes, \ie \ the NLO matrix element program or the \pythia \ Monte Carlo generation of LO matrix elements plus parton showers. The SCI model is, however, constructed as an add-on to a Monte Carlo and cannot be combined with the NLO program. We are therefore left with three combinations to consider. 

The first model we label {\bf CEM-NLO} and is the combination of the CEM model with the NLO program. This is done by implementing the formulas of section $2.2.1$ in the NLO program described in section $2.1$. In other words, we separate the colour singlet and colour octet contributions according to the CEM model and restrict the invariant mass region in order to produce charmonium or open charm. This gives CEM results accurate up to NLO in the pQCD part of the model.

The second model is {\bf CEM-PYTHIA}, where CEM has been implemented in \pythia \ version 5.7 \cite{Pythia}. The \ccbar \ pairs from the perturbative phase of \pythia \ are then treated with the CEM model with respect to their colour state and projected onto charmonium resonances. Here, one can exploit the advantages of the Monte Carlo approach applied to the CEM model. 

The third model, {\bf SCI-PYTHIA}, is to use the SCI model as implemented in \pythia \ 5.7 \cite{Pythia}. Since this model was introduced for other purposes, namely rapidity gaps in hard $ep$ and $p\bar{p}$ scattering, we are here testing its ability to describe different phenomena in a universal way.

As discussed, in the NLO-based model we use $m_c=1.5$~GeV and in the \pythia-based models we use $m_c=1.35$~GeV. These complete models also include a few other ingredients that need to be described. The pQCD parton level matrix elements must be folded with parton density distributions in the colliding hadrons in order to get the charm quark production cross section. To match the NLO program in CEM-NLO we use CTEQ5M \cite{CTEQ5M} for protons and SMRS \cite{SMRS} for pions, which are parameterisations in NLO based on the $\overline{MS}$ scheme. The LO pQCD formalism in \pythia \ is matched with LO parameterisations. For CEM-PYTHIA, we have used CTEQ4L \cite{CTEQ4L}, whereas for SCI-PYTHIA we used CTEQ3L \cite{CTEQ3L} in order to have exactly the same SCI model as compared to data in previous investigations. As parton parameterisations in the pion we have used LO GRV-P \cite{GRV-P} in both \pythia-based models. The uncertainty due to the choice of parton parameterisations has been investigated and found to be small, as will be discussed below. 

A further issue is the intrinsic transverse momentum $k_\perp$ of the partons in the colliding hadrons. Both in the NLO program and in \pythia , the components of a $\vec{k}_\perp$-vector can be chosen from a gaussian distribution. Its width is expected to be a few hundred MeV, corresponding to the Fermi motion within a hadron of size one fermi, but somewhat larger values are typically used phenomenologically. We use the width 0.6~GeV both in CEM-NLO and CEM-PYTHIA, whereas the slightly higher value of 0.8~GeV was used in SCI-PYTHIA. We note that this is in line with the new default value of 1~GeV in the latest \pythia \ version \cite{pythia6}. Variations of this gaussian width will be discussed below. 

The transverse momentum effect on the produced charmonium state due to the soft gluon exchanges that neutralize the colour of the $c\bar{c}$ pair has also been considered in the CEM-PYTHIA model. This was done by simply adding, to the charmonium momentum, a random \pt \ chosen from a gaussian distribution with default width 0.6~GeV, as used for the intrinsic transverse momentum. In the SCI model, the momentum of the soft exchanges has so far been neglected.

By comparisons of these three models we can separate different effects and gain understanding. With CEM implemented in the NLO program and in \pythia, we can explicitly compare the perturbative contributions, namely NLO versus LO plus the parton shower approximation of higher orders.
On the other hand, with SCI and CEM implemented in \pythia, we can explicitly compare these two non-perturbative models and see to what extent they can account for observed soft effects and thereby provide a better understanding of non-perturbative QCD. 

\section{Model results and experimental data}
Having discussed the details of the models in last section, we now turn to results of the models. Detailed comparisons between the models have been done as well as extensive comparison with data, both from fixed target experiments and the Tevatron collider. 

\subsection{Fixed target energies}
\label{fixten}
Here we compare the models to data from different experiments using beams of protons, antiprotons and pions at different energies in the range 125 to 800 GeV. The targets are different nuclei, but the experimental results are always rescaled to represent the cross section per nucleon. Based on this we compare directly with our models which do not include any nuclear effects but treat hadron-nucleon interactions. The relatively good agreement between models and data, suggests that there are no strong nuclear effects that have been missed. 

\begin{figure}[t]
\begin{center}
\centerline{\psfig{file=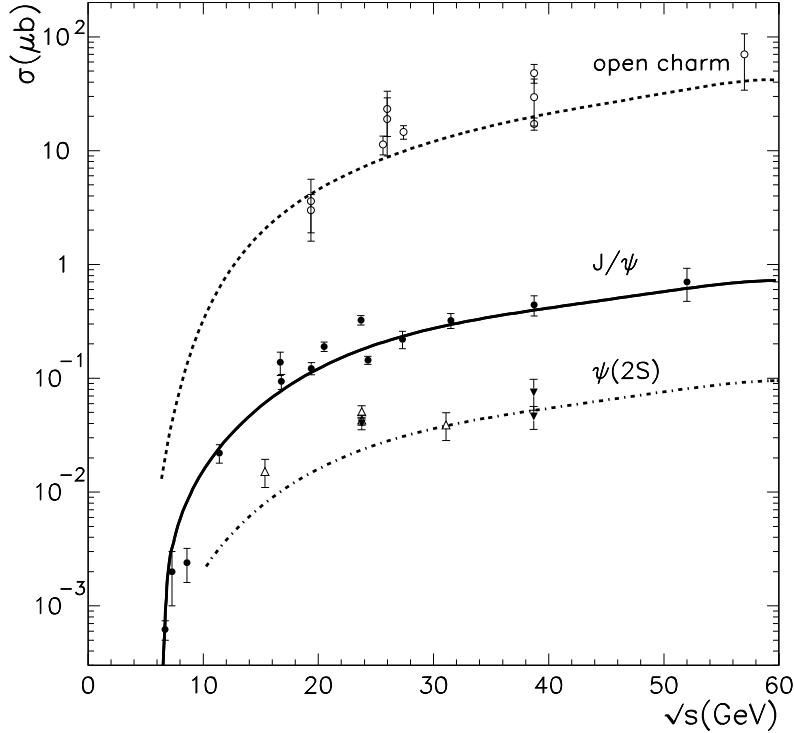,width=105mm}}
\caption{Total cross section of $J/\psi$, $\psi^{\prime}$ and open charm ($D$ mesons) in proton-proton interactions at invariant mass $\sqrt s$; curves obtained from the CEM-NLO model (colour evaporation model combined with NLO pQCD matrix elements) in comparison to a compilation of data \cite{sigtotpsi,sigtotop,pp800,300}.\label{sigtot}}
\end{center}
\end{figure}

As a first important overall test, we compare in Fig.~\ref{sigtot} the energy dependence of the cross section for $J/\psi$, $\psi^{\prime}$ and open charm obtained in the CEM-NLO model with available data. As can be seen, good agreement is obtained both for the overall normalization and the energy dependence. The CEM parameters were here chosen as $\rho_{J/\psi}=0.5$ and $\rho_{\psi^{\prime}}=0.066$. The former is consistent with earlier studies, whereas the latter is a new piece of information extracted here. The open charm cross section is obtained from all \ccbar \ in a colour octet state plus \ccbar \ in a singlet state with invariant mass above the $D\overline{D}$ threshold ({\it cf.} eq.~(\ref{2})). This gives a result in agreement with data without additional $K$-factors. Of course, the value of the charm quark mass does affect the overall normalization of the charm cross section, but we note that with $m_c=1.5$~GeV the CEM-NLO model gives the correct result throughout the whole study presented in this paper. Although the other models are not explicitly shown in Fig.~\ref{sigtot}, they can also reproduce these total cross sections as will be implicitly clear from the differential cross sections to which we now turn. 

\begin{figure}[t]
\begin{center}
\centerline{\psfig{figure=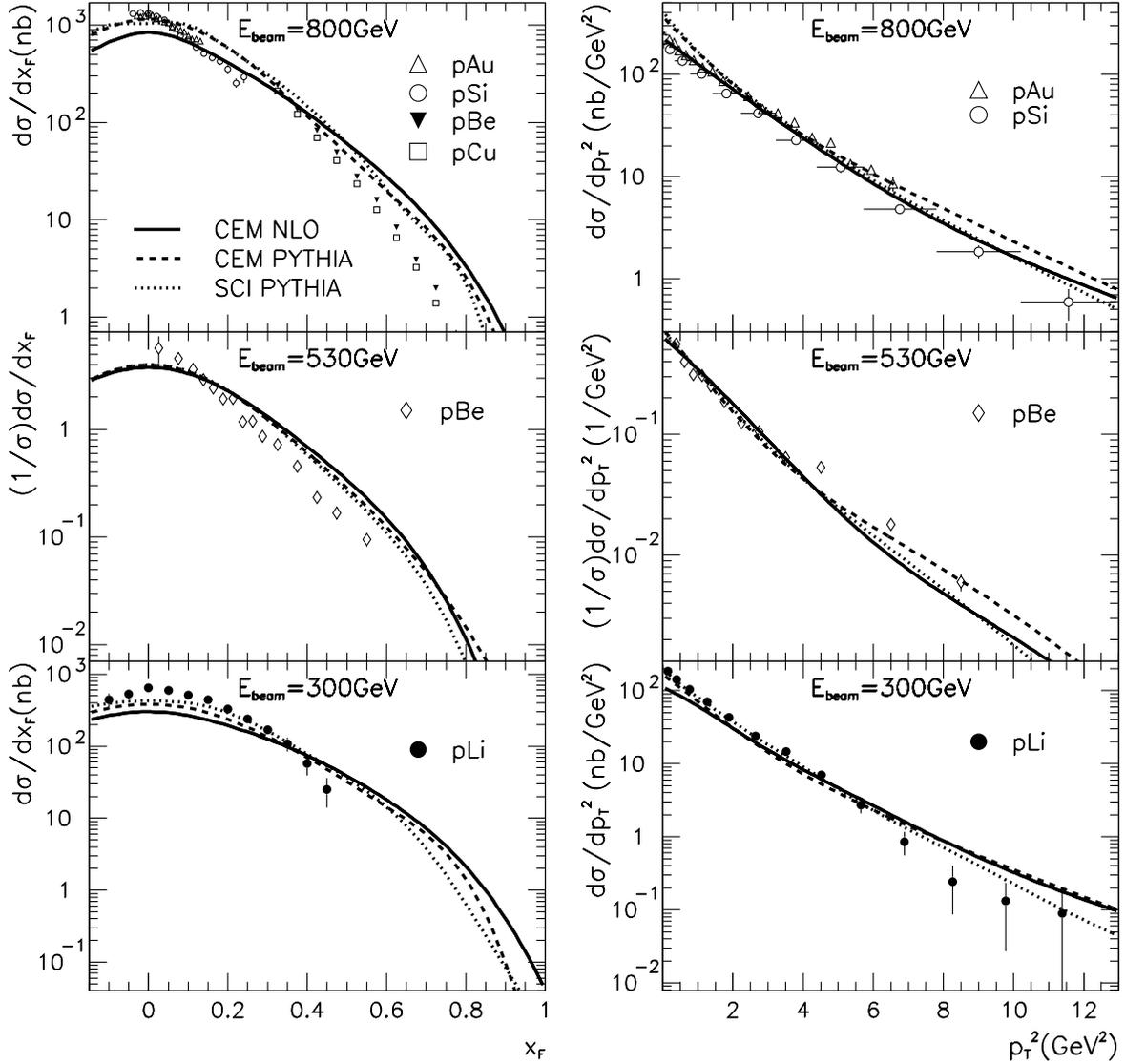,width=164mm}}
\caption{Distributions in $x_F$ and $p_\perp^2$ of $J/\psi$ produced with proton beams of energies $800$, $530$ and $300\, GeV$ on fixed target. Data \protect \cite{pp800,300,pp800k,pp530} compared to CEM based on NLO pQCD matrix elements, and CEM and SCI based on LO matrix elements plus parton showers in the \pythia \ Monte Carlo.\label{pp}}
\end{center}
\end{figure} 

More detailed information is available in differential cross sections. In particular, we will concentrate on distributions in Feynman-$x$, $x_F=p_\parallel/p_{\parallel max}$ in the hadronic cms, and transverse momentum, $p_\perp$, of the produced $J/\psi$. Fig.~\ref{pp} shows these distributions for proton beams of different energies. As can be seen, the data are approximately reproduced, both in shape and normalization, by all three models. Looking into the details of the \xf \ distributions, one can observe that all model curves, and in particular the \nlo, fall less steeply than the data and therefore overshoot somewhat at large \xf. The observed \pt distribution is better reproduced, with only small differences between the models. 

\begin{figure}[t] 
\begin{center}
\begin{tabular}{c c}
\psfig{file=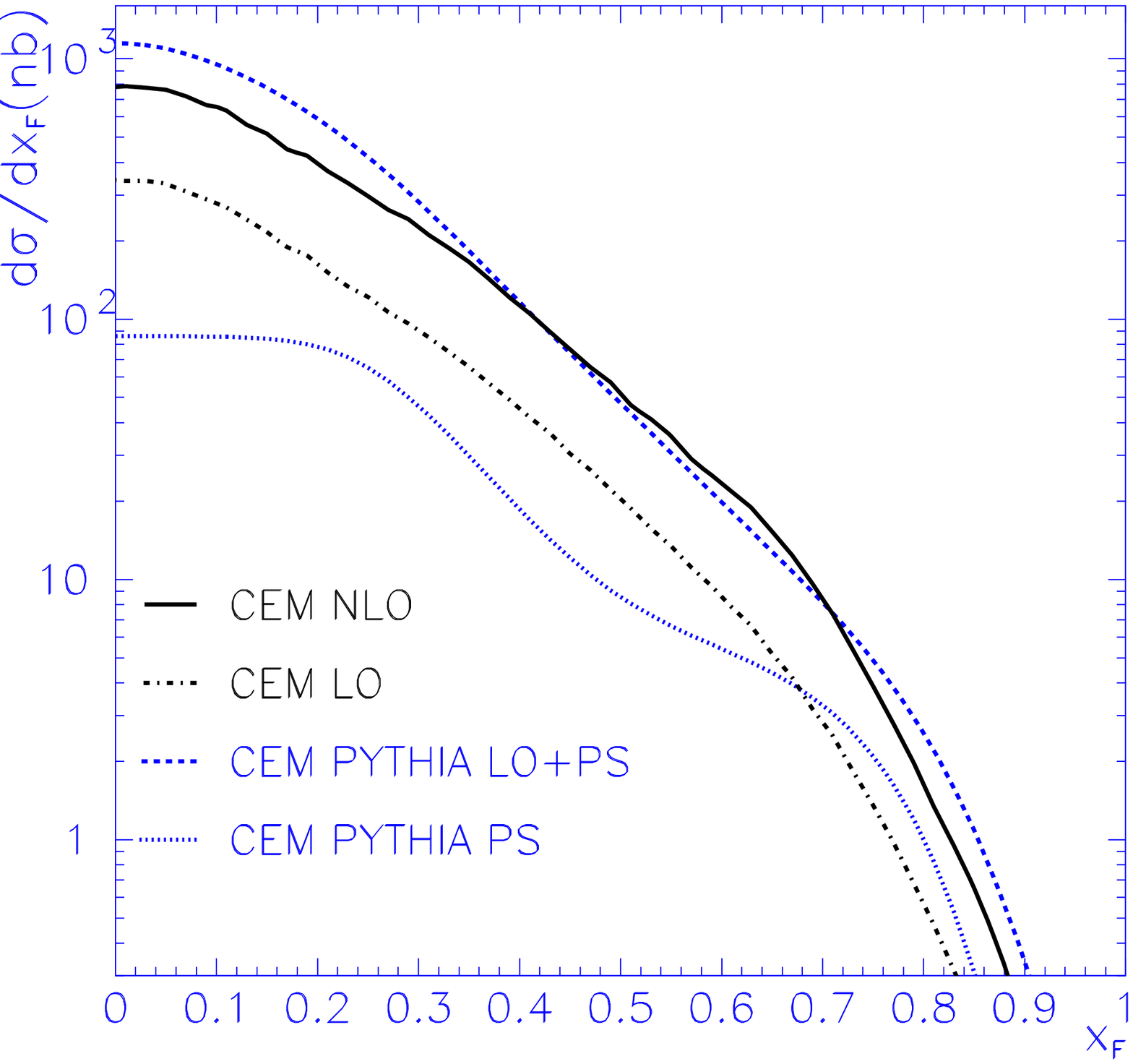,width=80mm} &
\psfig{file=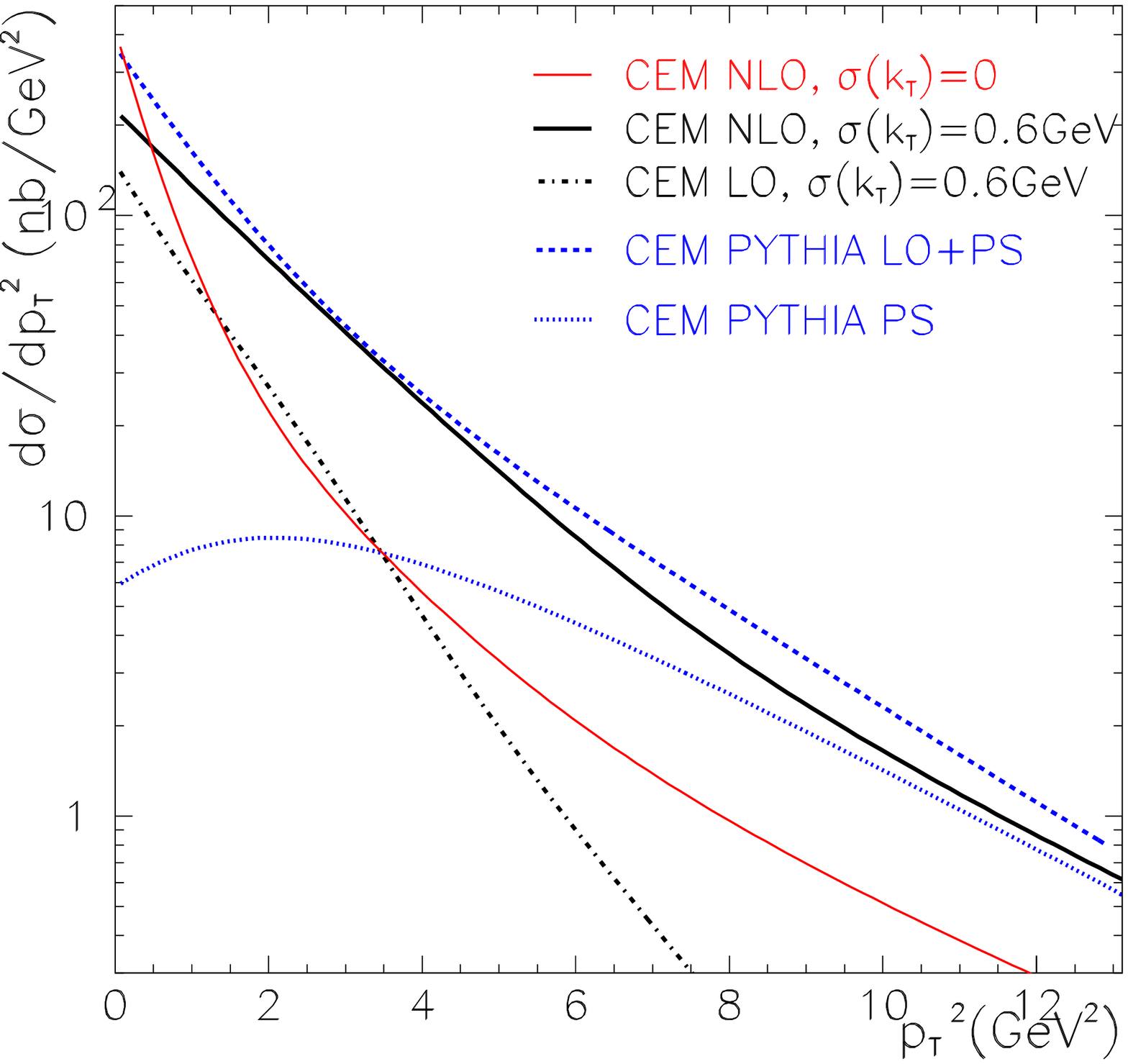,width=80mm}\\
\end{tabular}\vspace{-0.3cm}
\caption{Distributions in $x_F$ and $p_\perp^2$ of $J/\psi$ (in 800~GeV proton on proton as in Fig.~\ref{pp}) from variations of the pQCD treatment.\label{fig:xfteori} CEM based on the NLO program with $m_c=1.5$~GeV: NLO and LO matrix elements, NLO with no intrinsic $k_\perp$. CEM based on \pythia \ with $m_c=1.35$~GeV: LO matrix elements plus parton showers (PS) and PS contribution shown separately.}
\end{center}
\end{figure} 

Having CEM combined with different treatments of the pQCD production of \ccbar , we can now investigate pQCD effects in more detail on the \xf and \pt distributions. Fig.~\ref{fig:xfteori} illustrates this for the case of 800~GeV
proton energy. 

For the \xf distribution in Fig.~\ref{fig:xfteori}a, the result based on the full NLO program and that based on LO+PS agree quite well, except for a $20\%$ difference at small \xf. The LO contribution in the NLO calculation is less than 50\% \ of the total result, making the NLO corrections very important. In the LO+PS result, however, the PS contribution (with the \ccbar \ pair coming from initial and final state parton showers) gives a small contribution which is unimportant for the overall cross section. Thus, in the \pythia -based models, the \ccbar \ production is dominated by the LO $2\to 2$ processes having matrix elements with explicit charm quark mass. It is the use of a lower mass value, $m_c=1.35$~GeV, in \pythia \ which increases the cross section to become equal to the NLO one using the higher value $m_c=1.5$~GeV. We have cross-checked this within the NLO program, where the full result is essentially reproduced, both in shape and normalization, by the LO part if the lower mass $m_c=1.35$~GeV is used. This demonstrates that the NLO correction is essentially an overall $K$-factor from soft and virtual corrections. This $K$-factor can effectively be accounted for by using a lower charm quark mass in the LO matrix elements, which supports the phenomenological use of the  \pythia \ Monte Carlo. Although, the PS contribution is small in the region studied here, it can become important at higher cms energies as we shall see in Section 3.2.

For the understanding of the different pQCD contributions, the \pt distributions in Fig.~\ref{fig:xfteori}b is even more relevant. In the LO process the \ccbar \ pair cannot have a non-zero \pt , since there is no other parton to balance its \pt \ to obtain the essentially total zero \pt \ of the initial partons. Including the intrinsic $k_\perp$ of the initial partons, described with a gaussian distribution, gives rise to a steeply falling \pt distribution. The tail to large \pt \ observed in data (Fig.~\ref{pp}) can only be reproduced when higher order pQCD processes are included. The NLO program gives a \pt distribution with a much larger tail at large \pt, but it is still substantially affected by the inclusion of the intrinsic $k_\perp$ at the limited values of \pt accessible at fixed target energies. The \pt distribution resulting from the LO+PS in the \pythia \ approach, is at high-\pt dominated by 
\ccbar \ from gluon splittings in the partons showers, whereas the bulk of the cross section comes from the low-\pt region where the LO diagrams dominate. The total LO+PS result, which also includes a gaussian intrinsic $k_\perp$ with the same width 0.6~GeV, agrees quite well with the NLO result. This provides another comforting cross-check between the NLO and Monte Carlo approaches. 

Although Fig.~\ref{fig:xfteori} was for the case of 800 GeV proton beam energy, the main effects seen and the conclusions drawn are also valid for other energies and beam particles. We therefore do not show such detailed model decompositions for the following cases, but only compare the complete models with the data.

\begin{figure}[thb] 
\begin{center}
\centerline{\psfig{figure=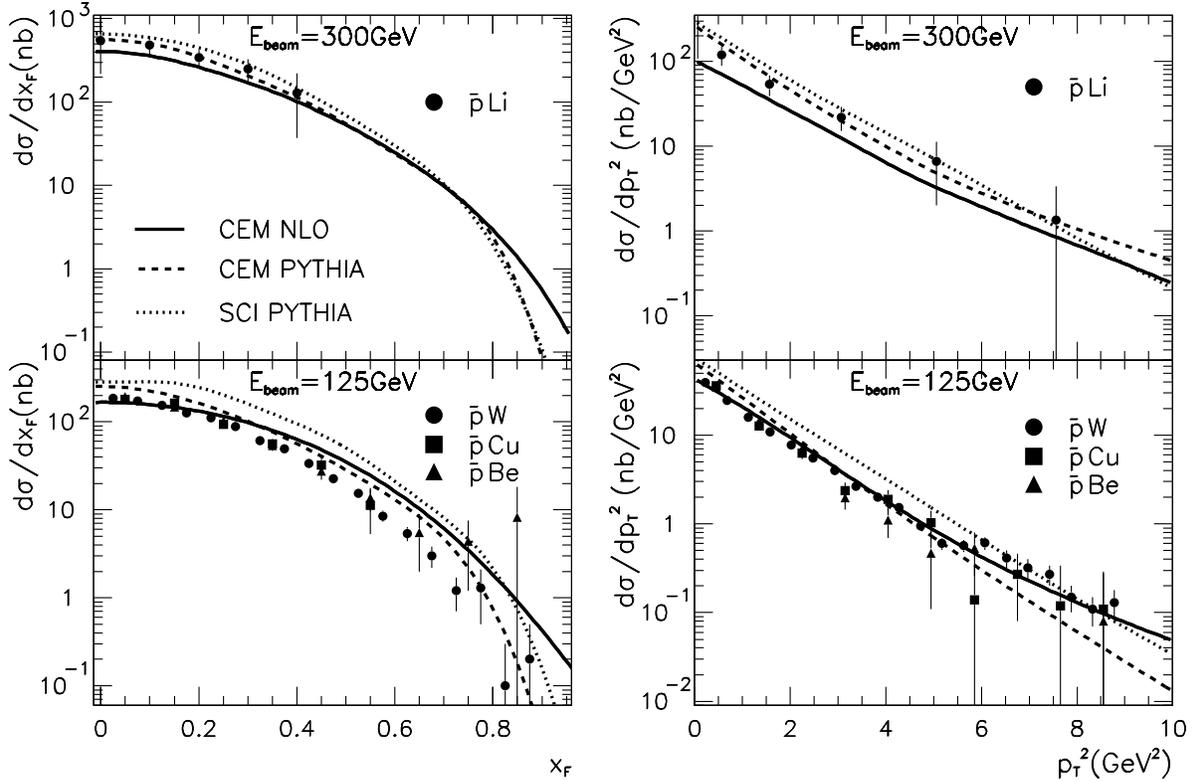,width=164mm}}
\vspace{-4.54cm}
\caption{\it Distributions in $x_F$ and $p_\perp^2$ of $J/\psi$ produced with antiproton beams of energies 300 and 125~GeV on fixed target. Data \protect \cite{300, 125} compared to CEM based on NLO pQCD matrix elements, and CEM and SCI based on LO matrix elements plus parton showers in the \pythia \ Monte Carlo.\label{pb}}
\end{center}
\end{figure} 

In Fig.~\ref{pb} we show the $x_F$ and $p_\perp$ distributions of \jpsi \ produced with antiproton beams of different energies. As can be seen, all three models reproduce the general features of the data, both in shape and normalization, although small deviations occur. In particular, \sci \ tends to overshoot the 125~GeV data. At this energy the observed \pt \ distribution tends to become flatter at high-\pt , which is well described by the \nlo \ model. 

\begin{figure}[t] 
\centerline{\psfig{figure=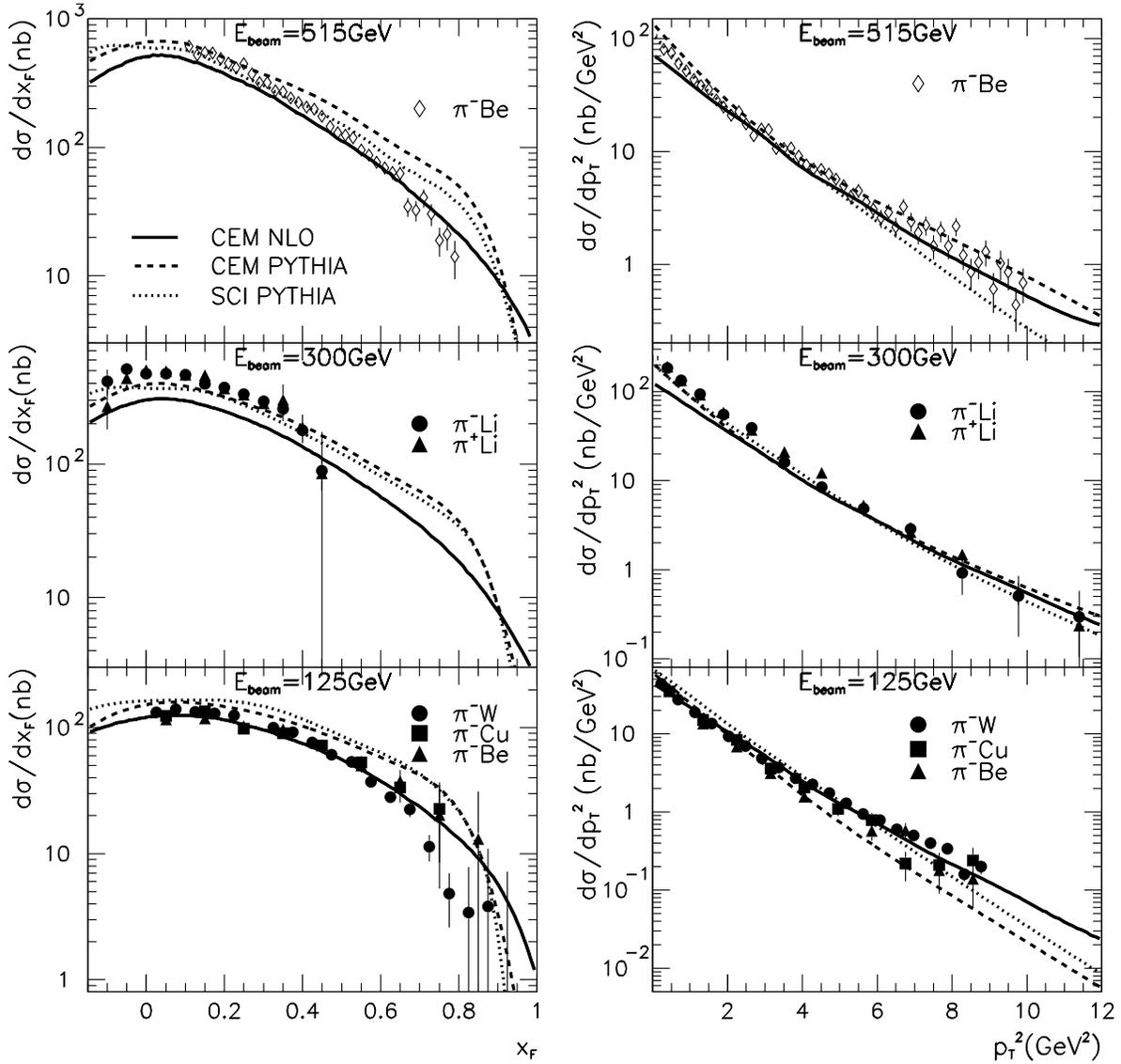,width=164mm}}
\caption{Distributions in $x_F$ and $p_\perp^2$ of $J/\psi$ produced with pion beams of energies of $515$, $300$ and $125\, GeV$ on fixed target. Data \protect \cite{pi515, 300, 125} compared to CEM based on NLO pQCD matrix elements, and CEM and SCI results based on LO matrix elements plus parton showers in the \pythia \ Monte Carlo.}
\vspace*{9mm}
\label{xfpi}
\end{figure} 

Fig.~\ref{xfpi} shows the corresponding results for pion beams of different energies. Again, the models give in general a reasonable description of the data. At 300~GeV the Monte Carlo models describe data better,
whereas at 125~GeV the \nlo \ works somewhat better.

This comparison with fixed target data on \jpsi \ production has shown that all three models reproduce the main features of the data. Looking into the details of the different distributions, one can certainly see discrepancies such that none of the models gives an excellent description of all data. This is expected in view of the simplicity of the non-perturbative parts of the models. The observed discrepancies may therefore indicate the need for improving the CEM or SCI models. One should also remember that nuclear effects were neglected, which may not be appropriate for a detailed understanding of the data. If one lets these three models illustrate the theoretical uncertainty, such that the three model curves represent an uncertainty band, then there are no really significant deviations from the data points in Figs.~\ref{pp}, \ref{pb} and \ref{xfpi}. This shows that the essential physical effects in \jpsi \ production at fixet target energies are accounted for by these model combining pQCD with soft colour exchanges. 

\begin{figure}[t] 
\begin{center}
\centerline{\psfig{figure=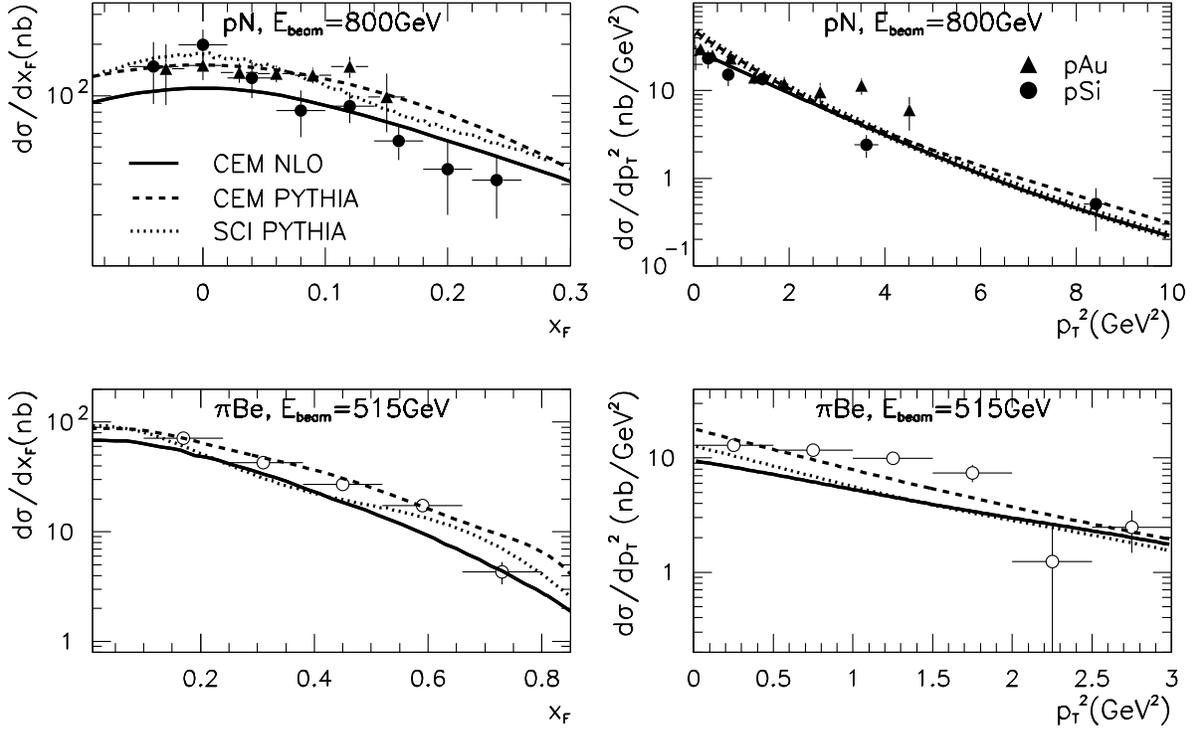,width=164mm}}
\vspace{-4.54cm}
\caption{Distributions in $x_F$ and $p_\perp^2$ of $\psi^{\prime}$ produced with proton and pion beams of energies $800$ and $515\, GeV$ on fixed target. Data \protect \cite{pp800,pi515} compared to CEM based on NLO pQCD matrix elements, and CEM and SCI based on LO matrix elements plus parton showers in the \pythia \ Monte Carlo. The normalization of CEM is given by $\rho_{\psi^{\prime}}=0.066$ and SCI is lowered by a factor four relative to spin statistics. \label{psi2s}}
\end{center}
\end{figure} 

Data on \psip \ production provide an additional testing ground for the models, which produce all charmonium states with the same dynamics. Although there are not as much data on \psip \ as on \jpsi \ production , one can still test the major features of the models. In Fig.~\ref{psi2s} we compare the models to data on the $x_F$ and $p_\perp$ distributions for \psip \ obtained with proton and pion beams. All models account quite well for the shape of the distributions. The proper normalization of CEM is obtained by chosing  $\rho_{\psi^{\prime}}=0.066$, \ie \ a factor 7.6 lower than $\rho_{J/\psi}$. The spin statistics used in SCI predicts only a factor two suppression of \psip , and must be lowered by an additional factor four in order to reproduce the data. Thus, although the spin statistics prescription was found to work at Tevatron energies \cite{sci-onium}, it fails for fixed target energies. This has prompted us to develop a more elaborate model for mapping \ccbar \ onto different charmonium resonances which will be presented in section 4. 

\subsection{Tevatron $p\bar{p}$ collider energy}
In order to investigate the energy dependence of the models further, we now compare them with data from $p\bar{p}$ collisions at $\sqrt{s}=1800$ GeV in the Fermilab Tevatron. The CDF~\cite{high-pt-data} and D0~\cite{D0} collaborations have data on the heavy quarkonium states \jpsi , \psip \ and \Y. These can be quite well reproduced by the SCI model \cite{sci-onium}. Here we make a similar comparison of the data with the results of the CEM. 

\begin{figure}[t]
\centerline{\psfig{file=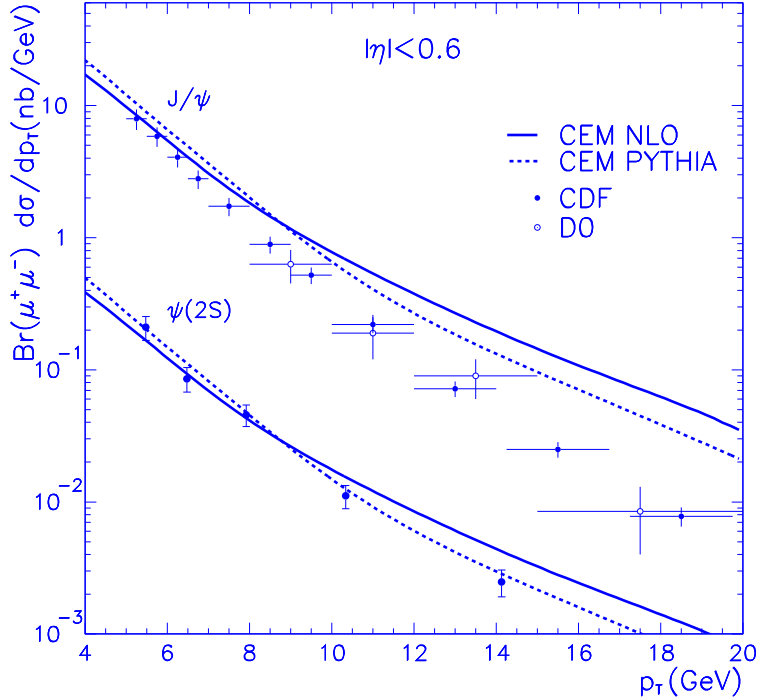,width=100mm}}
\caption{Distribution in transverse momentum of prompt \jpsi \ and \psip \ 
in $p\bar{p}$ interactions at $\sqrt{s}=1800$~GeV. Tevatron data from CDF~\cite{high-pt-data} and D0 \cite{D0} compared to results of CEM added to NLO matrix elements and to the \pythia \ Monte Carlo.}
\label{tevres}
\end{figure}

Fig.~\ref{tevres} shows that the CEM, both in its NLO and Monte Carlo version,
can reproduce the data. We have here used the same $\rho$-parameters as for the fixed target case above, although one could optimise the overall normalization to the Tevatron data by lowering them by 10-20\%. Such small adjustments are, however, not meaningful given the limited precision of the model. 

As discussed in Section 2, the LO processes for \ccbar \ production are unimportant for high-\pt \ charmonium. This is the case here, where the higher order pQCD processes give the main contribution, \ie \ order ${\cal O}(\alpha_s^3)$ tree diagrams in the NLO approach and the parton showers in the Monte Carlo approach.
The large $K$-factor associated with LO diagrams with soft and virtual corrections, may therefore be absent here. Indeed, the Monte Carlo approach does not need a lower charm quark mass to mimick $K$-factors, but the same value $m_c=1.5$~GeV is here used in both CEM-PYTHIA and CEM-NLO. The observed shapes of these \pt distributions are quite well represented by the models, although in particular CEM-NLO tends to give a too large cross section at the highest \pt. 

The results of an extrapolation of the models to LHC energies are presented in \cite{damet}, where also this prompt \jpsi \ production is considered as a background for CP-violation studies. 

\section{Improved mapping of \ccbar \ onto charmonium states}

As we have seen, the models considered give in general a good description of charmonium data both from fixed target and Tevatron energies. However, the overall energy-independent factors which describe the mapping of \ccb pair onto different charmonium states are not satisfactorily understood. In particular, spin statistics gives an acceptable ratio $\psi^\prime /\psi$ at the Tevatron~\cite{sci-onium}, but fails at fixed target energies where an extra suppression by a factor four is required. In other words, this ratio has an energy dependence which has not been accounted for. In this section we make an attempt to understand this problem by developing a model which introduces a correlation between the invariant mass of the \ccbar \ pair and the mass of the final charmonium state. 

The \ccbar \ pair is produced in a pQCD process with a continuous distribution of its invariant mass $m_{c\bar{c}}$. This has then to be mapped onto the discrete distribution of physical charmonium states. The soft interactions that turn the pair into a colour singlet and form the state, may very well change its mass by a few hundred MeV, which is the typical scale of the soft interactions, but larger mass shifts should be suppressed. The different charmonium states ($\eta_c$, \jpsi, $\chi_c$, \psip) are, however, separated in mass over a region of about 1~GeV. It therefore seems likely that the probability to form a particular charmonium state will depend on the original value of $m_{c\bar{c}}$ and not just on an overall factor, such as spin statistics. Thus, there should be a relatively larger probability to form states that are nearby in mass, than those further away on the mass scale. 
For example, a \ccbar \ with invariant mass just above the $2m_c$ threshold should have a larger probability to form a \jpsi \ than a \psip , and a \ccbar \ pair close to the open charm threshold should contribute more to \psip \ than to \jpsi . Based on these considerations we have constructed the following model. 

The smearing of the \ccbar \ mass due to soft interactions is described by the gaussian 
\begin{equation}
G_{sme}({\mccb}, m)= \exp\left( -\frac{(\mccb-m)^2}{2\sigma_{sme}^2}\right) ,
\label{gauss1}
\end{equation}
where the width $\sigma_{sme}$ should be a few hundred MeV. 
The different charmonium resonances $i$ are represented by functions $F_{i}(m_i, m)$ around their physical masses $m_i$. For a \ccbar \ pair with a given mass $m_{c\bar{c}}$, the probability to form a certain charmonium state is then 
\begin{equation}
{\cal {P}}_{i}({\mccb})= \frac {\int  G_{sme}({\mccb}, m) F_{i}(m_i, m) dm}
{\sum_{j} \int  G_{sme}({\mccb}, m) F_{j}(m_j, m) {dm}}
\approx 
\frac { s_i G_{sme}({\mccb}, m_i) }{\sum_{j} s_j G_{sme}({\mccb}, m_j) } , 
\label{Ps}
\end{equation}
where the sum is over all charmonium resonances. The approximate result, which is used in our practical calculations, is obtained by letting $F_{i}(m_i, m)=s_i\delta (m-m_i)$, \ie \ neglecting the width of the very narrow charmonium states and including the previously used spin statistical factor  $s_i=(2J_i+1)/n_i$. 

\begin{figure}[t]
\centerline{\psfig{file=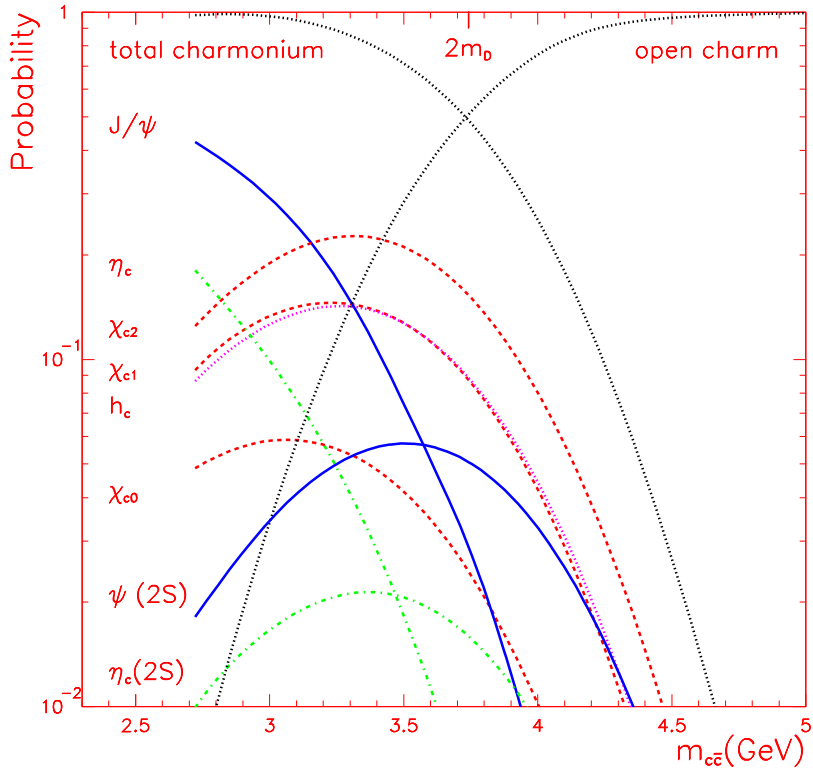,width=110mm}}
\caption{The probability functions $(1-A({\mccb})){\cal {P}}_{i}({\mccb})$ for the different charmonium states as obtained in the model using the smearing width $\sigma_{sme}=400\, MeV$. The resulting total probability for charmonium production and the remainder as open charm production are also shown.}
\label{dQ2PsDserr}
\end{figure}

With this smearing of the \ccbar \ invariant mass one must consider mass shifts
across the threshold for open charm $2m_D$. Thus, a pair originally below this border may, through the soft interactions, acquire a mass above this threshold and hence produce open charm instead. Conversely, the mass could be lowered such that a pair originally above this threshold fluctuates downwards into the charmonium region. For a given original $m_{c\bar{c}}$ value, the probability to end up in the region above $D\overline{D}$ threshold is given by the area of the smearing gaussian in that region, \ie \ 
\begin{equation}
A(m_{c\overline{c}})= \frac {1}{\sqrt{2\pi}\sigma_{sme}} \int_{2m_D}^{\infty} \exp\left( -\frac{(\mccb-m)^2}{2\sigma_{sme}^2}\right) dm 
=  \frac{1}{2} {\rm erfc}\left( \frac{2m_D - \mccb}{\sqrt{2\pi}\sigma_{sme}}\right) \,,
\end{equation}
where erfc(x) is the complementary error function. The complementary area $(1-A)$, multiplied with the function ${\cal{P}}_i$, will then give the probability to produce a given charmonium state. These probabilities for the different charmonium states are shown in Fig.~\ref{dQ2PsDserr}. For a given original \ccbar \ mass $m_{c\bar{c}}$ one can here see the fractional production of different states. The sum of all charmonium states is also shown together with the remainder giving open charm. One notices the non-zero contributions for charmonium also in the region above the $D\overline{D}$ threshold as well as some open charm production for $m_{c\bar{c}}$ originally below this threshold. 

By folding these charmonium probability functions with the distribution in $m_{c\bar{c}}$ obtained from pQCD, \ie \ ${d\sigma}_{c\overline{c}} /dm_{c\overline{c}}$, the cross section for a given charmonium state is obtained as 
\begin{equation}
\sigma_{i}= \int_{2m_c}^{\sqrt{s}} dm_{c\overline{c}} 
\frac{{d\sigma}_{c\overline{c}}} {dm_{c\overline{c}}}(1-A(\mccb)){\cal {P}}_{i}({\mccb})\, .
\end{equation}

\begin{figure}[p]
\centerline{\psfig{file=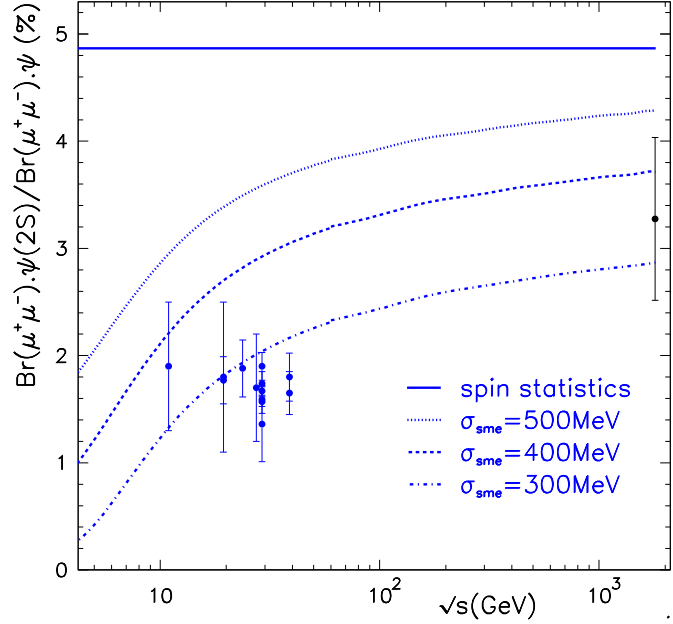,width=90mm}}
\caption{The ratio of \psip \ to \jpsi \ production (times their branching ratios for decay into $\mu^+\mu^-$) in hadron-hadron interactions of cms energy $\sqrt{s}$. Data \cite{Na38p450,high-pt-data} compared to simple spin statistics and to our model with different gaussian smearing widths applied to CEM.}
\label{dQ2gaussbrrates}
\end{figure}

One can apply this mapping procedure to the CEM and SCI models and compare with data. In Fig.~\ref{dQ2gaussbrrates} we show the ratio of \psip \ to \jpsi \ production as a function of the cms energy of the hadron-hadron interaction. As can be seen, simple spin statistics cannot describe this ratio, whereas our elaborated model can. In particular, it gives a characteristic energy dependence of the kind indicated by the data. 
\begin{figure}[p]
\centerline{\psfig{file=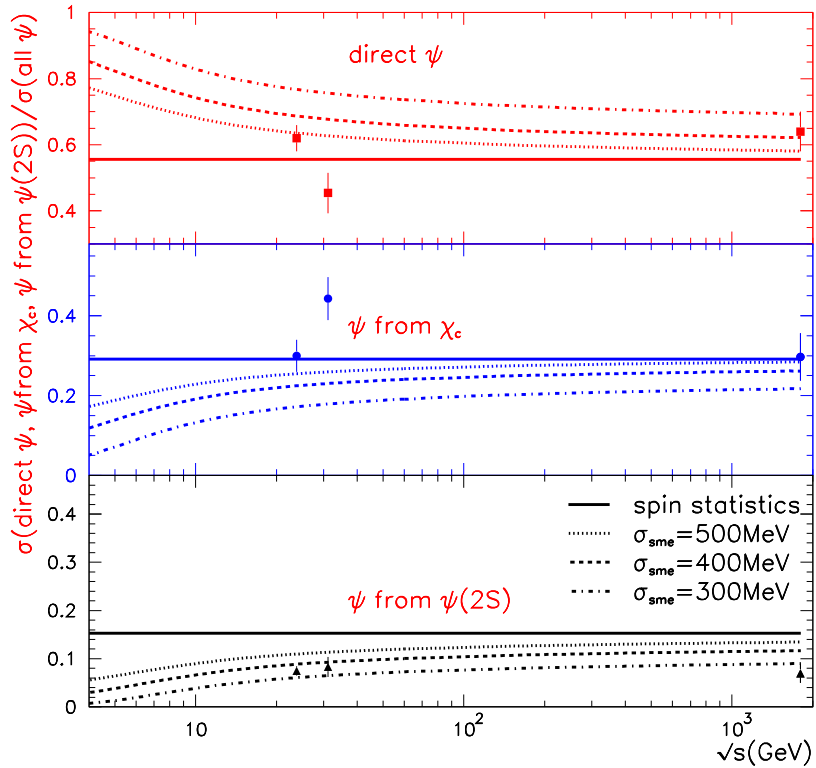,width=110mm}}
\caption{Fractions of \jpsi \ produced directly, and coming from the decay of $\chi_c$ and $\psi^{\prime}$ states. Data \cite{ppi300E705, pi515koreshev, high-pt-data} compared to simple spin statistics and to our model with different gaussian smearing widths.}
\label{dQ2gaussfract}
\end{figure}

The fractions of \jpsi \ produced directly, coming from decays of $\chi_c$ states and from $\psi^{\prime}$ are shown in Fig.~\ref{dQ2gaussfract}. 
Again, the simple spin statistical factors cannot give a good representation of all data, whereas our elaborated model gives an improved description of data without strong disagreements. 

The ratio $\chi_{c1}/\chi_{c2}$ is $3:5$ according to spin statistics. Since these $\chi_c$ states are so close in mass, the mass correlation introduced here does not change the spin statistics result significantly. Furthermore, the experimental situation is very unclear for this ratio. The latest value from E771 is $0.53\pm 0.20\pm 0.07$ and they give a world average of $0.31\pm 0.14$ for fixed target experiments \cite{E771}. In $p\bar{p}$ at the Tevatron, on the other hand, one has found an approximately equal production of these two states \cite{tev2001}. 

We have checked that the results for \jpsi \ presented in Section 3 do not change significantly if one uses this improved mapping procedure instead of spin statistics. The difference is less than $20\%$, which is within the uncertainty of the models.

Considering all observables from Fig.~\ref{dQ2gaussbrrates} and \ref{dQ2gaussfract}, the preferred value for the $\sigma_{sme}$ is 400~MeV or lower. This fits well with the expectation for the non-perturbative dynamics that this model should describe. The essential new ingredient of this model, namely introducing a correlation between the invariant mass of the \ccbar \ pair and the masses of the different charmonium states, is found to give an improved description of the ratios of different charmonium states as compared to using constant factors based on spin statistics. In particular, these ratios acquire an energy dependence which is particularly strong at low energies where threshold effects are more pronounced. A tendency for such an effect is observed in data, but further verification is desirable. 

\section{Conclusions}

This comprehensive study of charmonium production in hadronic interactions has demonstrated the importance of both hard, perturbative and soft, non-perturbative QCD dynamics. For the pQCD production of a \ccbar \ pair, we have shown the importance of higher order contributions. Using a full NLO matrix element program, we find a large increase (factor two) of the total cross section. Since this comes mainly from soft and collinear gluon emissions combined with virtual corrections, this can be effectively accounted for by an 
overall $K$-factor. Because of this, the NLO results can be reproduced with the leading order matrix elements having a reduced charm quark mass to increase the cross section correspondingly. In this way the phenomenologically more useful \pythia \ Monte Carlo can be used for practical purposes. The tail of the cross section at high \pt \ of the \ccbar \ pair is, however, dominated by higher order tree diagrams available as ${\cal O}(\alpha_s^3)$ in the NLO program and in the parton showers of the Monte Carlo approach. 

The formation of charmonium states from a \ccbar \ pair is described by
non-perturbative dynamics. In order to understand the overall magnitude of
charmonium production, one must here take into account that a fraction of the
more abundantly produced \ccbar \ pairs in a colour octet state are turned into
a colour singlet state. This is the essence of the Colour Evaporation Model and the Soft Colour Interaction model considered in this study. The CEM uses simple SU(3) colour factors to give a random colour charge to a \ccbar \ pair. For colour singlets with mass below the open charm threshold, different charmonium states are formed with probabilities given by phenomenological parameters ($\rho_{J/\psi}$ etc). SCI is a Monte Carlo model which simulates soft gluon exchange between partons and remnants emerging from a hard scattering. The resulting colour singlet \ccbar \ pairs (below open charm threshold) form different charmonium states based on spin statistics. 

Adding CEM or SCI on the pQCD treatment provides complete models that are able to describe the observed charmonium production. The total cross section as well as the $x_F$ and \pt \ distributions for \jpsi \ produced with different beam particles and targets can be well described over the large energy range provided by fixed target experiments and the Tevatron collider. When using the full NLO matrix elements, there is no need to introduce an arbitrary $K$-factor since the correct normalization is obtained with the CEM using conventional parameter values. The SCI model, which is absolutely normalised by determining its single  parameter to fit rapidity gap data, gives automatically the correct normalization. 

Looking into the details of the $x_F$ and \pt \ distributions, some deviations of the model results from the data points can be observed. However, letting the different models represent the theoretical uncertainty, then there are no significant deviations. This shows that the essential physical effects in \jpsi \ production are accounted for in this kind of soft colour exchange models. 

The models were also tested with respect to the production of other charmonium states, although data is here much more scarce. The $x_F$ and \pt \ distributions available for \psip \ can be reproduced by CEM when the free  normalization parameter $\rho_{\psi^\prime}$ is fitted. The spin statistics for the different charmonium states used in the SCI model, works reasonably for \psip \ data at the Tevatron, but gives a factor four too high \psip \ cross section at fixed target energies. Such an energy dependence of the relative rates of different charmonium states requires additional features of the models.

We have made an attempt to solve this problem by a more elaborate model to map the \ccbar \ pairs onto the physical charmonium resonances. The probability to form a certain charmonium state is then not only given by spin statistics, but also by a correlation between the \ccbar \ invariant mass and the charmonium mass. Resonances with mass closer to the original \ccbar \ are thereby favoured and fluctuations across the open charm threshold are considered. This improves the description of the relative rates of different charmonium states with an energy dependence as indicated by data. 

Our study has demonstrated that the main features of hadroproduction of charmonium can be described in these models combining pQCD and effects of soft colour exchanges. This shows, in particular, that these models for the soft QCD dynamics contain the essential effects and therefore improve our understanding of non-perturbative QCD. 

{\bf Acknowledgments:} We are grateful to Johan Rathsman for helpful discussions. One of us (C.B.M.) would like to thank the High Energy Physics group in Uppsala for the kind hospitality. This work was partially financed by Conselho Nacional de Desenvolvimento Cient\'{\i}fico e Tecnol\'ogico (CNPq) and Funda\c{c}\~ao Coordena\c{c}\~ao de Aperfei\c{c}oamento de Pessoal de N\'{\i}vel Superior (CAPES), Brazil, and by the Swedish Natural Science Research Council.

\end{document}